\documentclass[aps,twocolumn,floatfix,groupedaddress,nofootinbib]{revtex4}
\usepackage{csquotes}
\usepackage{graphicx,color}
\usepackage{float}
\usepackage[caption=false]{subfig}
\usepackage{amsmath}
\usepackage{amsmath}
\usepackage{multirow}
\usepackage{dsfont}

\begin{document}

\title{Isomorph-based empirically modified hypernetted-chain approach for strongly coupled Yukawa one-component plasmas}

\author{P. Tolias and F. Lucco Castello}
\affiliation{Space and Plasma Physics, Royal Institute of Technology, Stockholm, SE-100 44, Sweden}

\begin{abstract}
\noindent Isomorph theory is employed in order to establish a mapping between the bridge function of Coulomb and Yukawa one-component plasmas. Within an exact invariance ansatz for the bridge functions and by capitalizing on the availability of simulation-extracted Coulomb bridge functions, an analytical Yukawa bridge function is derived which is inserted into the integral theory framework. In spite of its simplicity and computational speed, the proposed integral approach exhibits an excellent agreement with computer simulations of dense Yukawa liquids without invoking adjustable parameters.
\end{abstract}

\maketitle

\section{Introduction}\label{intro}

\noindent Yukawa one-component plasmas (YOCP) are pair additive systems of charged point particles interacting via the potential $u(r)=(Q^2/r)\exp{\left(-r/\lambda\right)}$ where $Q$ is the particle charge and $\lambda$ is the screening length resulting from the polarizable neutralizing background. The thermodynamic state of the three-dimensional YOCP is completely specified by two independent dimensionless variables, the coupling parameter $\Gamma$ and the normalized screening parameter $\kappa$ defined by\,\cite{intro00,intro01,intro02,intro03}
\begin{equation*}
\Gamma=\beta\frac{Q^2}{d}\,,\,\,\,\,\kappa=\frac{d}{\lambda}\,,\label{basic}
\end{equation*}
where $d=\left(4\pi{n}/3\right)^{-1/3}$ is the Wigner-Seitz or ion-sphere radius, $n$ is the number density, $\beta=1/(k_{\mathrm{b}}T)$, $T$ is the temperature and $k_{\mathrm{b}}$ is the Boltzmann constant. On the contrary, one-component plasmas (OCP) are characterized by a rigid neutralizing background resulting in unscreened bare Coulomb interactions ($\lambda\to\infty$, $\kappa=0$) and their phase diagram is described solely by one dimensionless state variable, the coupling parameter $\Gamma$.

In the $\Gamma\ll1$ range the equilibrium YOCP behaves as a non-interacting gas, in the range $\Gamma\gg1$ it is crystallized (bcc or fcc) and for intermediate $\Gamma$ it behaves as a dense liquid. The YOCP paradigm has been long studied in statistical mechanics being a model system whose interactions can vary from extremely short range hard sphere-like ($\kappa\to\infty$) to infinitely long range Coulomb-like ($\kappa\to0$) depending on the phase diagram region\,\cite{intro00}. The experimental realization of the strongly coupled YOCP in complex (dusty) plasmas\,\cite{intro01,intro02,intro03}, charge-stabilized colloidal suspensions\,\cite{intro04,intro05}, ultra-cold neutral plasmas\,\cite{intro06,intro07} and their relevance to warm dense matter\,\cite{intro08,intro09} has provided significant impetus to such studies. In fact, a large number of theoretical, computational and experimental works have focused on the structure and thermodynamics\,\cite{intro11,intro12,intro13}, transport coefficients and collective modes\,\cite{intro14,intro15,intro16} as well as the phase behavior of the YOCP\,\cite{intro17,intro18}.

On the computational front, the static properties of the three-dimensional YOCP have been systematically studied with Molecular Dynamics (MD)\,\cite{Hamagu1,Hamagu2,Hamagu3}, Langevin Dynamics (LD)\,\cite{Bonitz1,Bonitz2} and Monte Carlo (MC) simulations\,\cite{Meijer1,Cailol1,Cailol2}. As a consequence, simple analytical expressions have been proposed that allow for an accurate estimation of fundamental thermodynamic properties\,\cite{practi1} and even of reduced particle distribution functions\,\cite{practi2}. On the theoretical front, the static YOCP characteristics have been investigated with various integral theory methods including the hypernetted-chain\,\cite{WittHNC}, the soft mean spherical\,\cite{SMSAme1,SMSAme2}, the discretized Rogers-Young\,\cite{DRYpape}, the variational modified hypernetted-chain\,\cite{VMHNCre} and the empirically modified hypernetted-chain approximations\,\cite{EMHNCre} as well as with semi-phenomenological approaches such as the Debye-H\"uckel plus hole\,\cite{Khrapa1} and ion sphere models\,\cite{Khrapa2}. It is also worth pointing out that static properties constitute necessary input for advanced theoretical descriptions of dynamic characteristics; for instance the quasi-localized charge approximation of collective modes requires knowledge of the pair correlation function\,\cite{dynami1}, whereas the viscoelastic-dynamic density functional theory of wave dispersion\,\cite{dynami2}, the non perturbative model-free moment approach of dynamic density-density correlations\,\cite{dynami3} and the mode coupling theory of the glass transition\,\cite{dynami4} require knowledge of the structure factor.

In the present work, we formulate a new integral theory method for the study of the YOCP structural and thermodynamic properties. The YOCP bridge function is obtained from the simulation-extracted OCP bridge function\,\cite{OCPbri1} by taking advantage of recent theoretical developments on the approximate \enquote{hidden} scale invariance of strongly coupled systems\,\cite{Roskil0}. Within the ansatz that the YOCP bridge function remains constant while traversing any isomorph curve, this task can be achieved by utilizing a closed-form mapping that extends the OCP to the entire dense fluid $(\Gamma,\kappa)$ phase diagram through approximate configurationally adiabatic paths. The insertion of the derived YOCP bridge function into the standard integral theory framework results to a fast and accurate method for the computation of static correlations. The proposed isomorph-based empirically modified hypernetted-chain approximation exhibits excellent agreement with computer simulations of dense YOCP liquids. Despite the fact that this approach does not involve tunable or adjustable parameters, it performs consistently better than other rigorous integral theory approximations previously employed for the YOCP\,\cite{WittHNC,SMSAme1,SMSAme2,DRYpape,VMHNCre,EMHNCre} as far as thermodynamic properties and structural quantities are concerned. The generalization of the proposed approach to a broader class of strongly coupled systems beyond the YOCP is discussed and possible further improvements are suggested.

\section{Theory}\label{theoretical}

\noindent In the integral theory of liquid structure, the central relation is the well-known Ornstein-Zernike (OZ) equation which has the convolution form\,\cite{bookre1,bookre2}
\begin{equation}
h(r)=c(r)+n\int\,c(r^{\prime})h(|\boldsymbol{r}-\boldsymbol{r}^{\prime}|)d^3r^{\prime}\,,\label{OZequation}
\end{equation}
where $h(r)=g(r)-1$ is the total correlation function, $g(r)$ is the pair correlation or radial distribution function and $c(r)$ is the direct correlation function. The equation essentially serves as the definition of $c(r)$ and thus an additional expression is necessary. The OZ closure equation is derived from a cluster diagram analysis and reads as\,\cite{bookre1,bookre2}
\begin{equation}
g(r)=\exp{\left[-\beta{u}(r)+h(r)-c(r)+B(r)\right]}\,,\label{OZclosure}
\end{equation}
where $B(r)$ is known as bridge function. The latter equation is also formally exact, in the sense that $B(r)$ has a well-defined diagrammatic representation. Nevertheless, it is not possible to calculate the resulting infinite series, hence this equation essentially serves as the definition of $B(r)$\,\cite{bookre3} and yet another expression is needed. In fact, the distinguishing characteristic among different integral theory approaches solely stems from the approximate expression assumed for the connection between $B(r)$ and $g(r),\,c(r)$\,\cite{bookre1,bookre2,bookre3}.

The most fundamental bridge function closure schemes are the following; $B(r)=0$ within the hypernetted-chain approximation (HNC), $B(r)=\ln{\left[1+\gamma(r)\right]}-\gamma(r)$ within the Percus-Yevick approximation (PY) and $B(r)=\ln{\left[g(r)\right]}-g(r)+1$ within the asymptotic branch of the soft mean spherical approximation (SMSA)\,\cite{bookre1,bookre2,bookre3}. Here the indirect correlation function $\gamma(r)=h(r)-c(r)$ has been introduced for convenience. It is worth pointing out that several approaches have been developed that either insert adjustable parameters to the aforementioned closure schemes or interpolate between different approximations in a manner that enforces thermodynamic consistency, see for instance Ref.\cite{approx6}.

Finally, we note that the above approaches introduce $B[g,c]$ as a functional of the pair and direct correlation functions, which implies their applicability for arbitrary pair potentials. There are also approaches that directly introduce $B(r)$ as a function of the distance which are typically referred to as modified hypernetted-chain approximations (MHNC). It is evident that, in case such approaches do not contain adjustable parameters, their applicability should be limited to a specific type of interaction. The approach described herein belongs to the latter category. However, it can be straightforwardly generalized for a certain class of strongly coupled liquids provided that certain simulation input is available a priori.

\subsection{Isomorph theory and Roskilde-simple systems}\label{theoryRoskilde}

\noindent The so-called Roskilde-simple (or for brevity R-simple) systems exhibit strong correlations between their virial ($W$) and their potential energy ($U$) equilibrium fluctuations\,\cite{Roskil0,Roskil1,Roskil2,Roskil3,Roskil4,RoskilN}. This class of condensed matter (dense liquids or solids) is practically defined by possessing correlation coefficients $R=\langle\Delta{U}\Delta{W}\rangle/\sqrt{\langle(\Delta{U})^2\rangle\langle(\Delta{W})^2\rangle}$ that satisfy $R\gtrsim0.9$ within an extended region of their phase diagram\,\cite{Roskil1,Roskil3}. In the above, the bracket operator denotes statistical averaging in the canonical ensemble and the $\Delta$ operator denotes the deviation from the thermodynamic mean, \emph{i.e.} $\Delta{A}=A-\langle{A}\rangle$. The strong correlation property is equivalent to the existence of isomorph curves; phase diagram lines of constant excess entropy along which the structure and dynamics in properly reduced units are approximately invariant\,\cite{Roskil0,Roskil4,RoskilN}. Essentially, the isomorph mapping roughly transforms the phase diagram of a one-component system from two-dimensional to one-dimensional. The mapping is exact only for systems that are characterized by $R\equiv1$, which mainly corresponds to the long known case of inverse power law pair potentials\,\cite{Roskil5}.

In a recent investigation\,\cite{Roskil6}, it was demonstrated that the strongly coupled YOCP system is R-simple. In particular, MD simulations revealed exceptionally high values of $R>0.99$ at all the tested state points spanning a large portion of the liquid phase. This is a rather anticipated outcome in view of the facts that the YOCP abides by Rosenfeld's excess entropy and melting temperature scalings of transport coefficients\,\cite{scalin1,scalin2,scalin3}, obeys the Rosenfeld-Tarazona scaling for the thermal correction to the fluid Madelung excess internal energy\,\cite{scalin4,scalin5} as well as is characterized by melting and glass transition lines of a similar functional form\,\cite{scalin6}. In the same investigation\,\cite{Roskil6}, it was also identified that the isomorph curve is successfully described by
\begin{equation}
\Gamma_{\mathrm{iso}}(\Gamma,\kappa)=\Gamma{e}^{-\alpha\kappa}\left[1+(\alpha\kappa)+\frac{1}{2}(\alpha\kappa)^2\right]=\mathrm{const.}\,,\label{eq-mapping}
\end{equation}
where $\alpha=\Delta/d=\left(4\pi/3\right)^{1/3}$ is the ratio of the cubic mean inter-particle distance $\Delta=n^{-1/3}$ over the Wigner-Seitz radius $d=\left(4\pi{n}/3\right)^{-1/3}$.

We point out that Vaulina and Khrapak originally proposed this analytical expression in order to empirically describe the YOCP melting line as revealed by MD simulations\,\cite{Hamagu1,Hamagu2,Hamagu3} but also presented a heuristic derivation by applying Lindemann's melting rule and employing the characteristic dust lattice wave frequency in the determination of the mean squared displacement\,\cite{isoKhr1,isoKhr2}.

\subsection{Bridge function in the OCP limit}\label{theoryBridgeOCP}

\noindent The most accurate extraction of the OCP bridge function from computer simulations was carried out by Ichimaru and collaborators\,\cite{OCPbri1}. Before proceeding, it is necessary to introduce the potential of mean force $w(r)$ defined by $g(r)=\exp{[-\beta{w}(r)]}$ and the screening potential $H(r)$ defined through the decomposition $w(r)=u(r)-H(r)$. Combining with the exact OZ closure, the definitions directly lead to $B(r)=\beta{H}(r)-h(r)+c(r)$ which can be used for the bridge function determination\,\cite{OCPbri2}.

The bridge function extraction from MC data necessarily involves short- and long-range extrapolations\,\cite{OCPbri1}. Ichimaru's short-range extrapolation was based on truncating the exact Widom expansion for $\beta{H}(r)$ after the second term\,\cite{OCPbri3} and the long-range extrapolation was based on the compressibility sum rule for the structure factor\,\cite{OCPbri4}. It is worth noting that the premature truncation of the $\beta{H}(r)$ power series was strongly criticized by Rosenfeld\,\cite{OCPbri5,OCPbri6}. The parameterization reads as\,\cite{OCPbri1}
\begin{align}
&B_{\mathrm{OCP}}(r,\Gamma)=\Gamma\left[-b_0(\Gamma)+c_1(\Gamma)\left(\frac{r}{d}\right)^4+c_2(\Gamma)\left(\frac{r}{d}\right)^6\right.\nonumber\\&\,\qquad\qquad\qquad\left.+c_3(\Gamma)\left(\frac{r}{d}\right)^8\right]\exp{\left[-\frac{b_1(\Gamma)}{b_0(\Gamma)}\left(\frac{r}{d}\right)^2\right]}\,,\label{eq-bridge1}\\
&b_0(\Gamma)=0.258-0.0612\ln{\Gamma}+0.0123(\ln{\Gamma})^2-\frac{1}{\Gamma}\,,\label{eq-bridge2}\\
&b_1(\Gamma)=0.0269+0.0318\ln{\Gamma}+0.00814(\ln{\Gamma})^2\,,\label{eq-bridge3}\\
&c_1(\Gamma)=0.498-0.280\ln{\Gamma}+0.0294(\ln{\Gamma})^2\,,\label{eq-bridge4}\\
&c_2(\Gamma)=-0.412+0.219\ln{\Gamma}-0.0251(\ln{\Gamma})^2\,,\label{eq-bridge5}\\
&c_3(\Gamma)=0.0988-0.0534\ln{\Gamma}+0.00682(\ln{\Gamma})^2\,.\label{eq-bridge6}
\end{align}
The validity range of Eqs.(\ref{eq-bridge1},\ref{eq-bridge2},\ref{eq-bridge3},\ref{eq-bridge4},\ref{eq-bridge5},\ref{eq-bridge6}) was specified to be $5<\Gamma\leq180$, even though MC simulations were carried out only for four values of the OCP coupling parameter, namely $\Gamma=10,\,40,\,80$ and $160$\,\cite{OCPbri1,OCPbri2}.

The upper validity threshold of $\Gamma_{\mathrm{th}}^{\mathrm{u}}=180$ simply corresponds to the coupling parameter at the bcc freezing transition, as determined by early computer simulations\,\cite{OCPmold}. We point out that recent MD works resulted in $\Gamma_{\mathrm{m}}^{\mathrm{OCP}}=171.8$\,\cite{Hamagu2}, which can be considered as a more appropriate threshold. Further extrapolations in the supercooled liquid regime may introduce errors, since neglected $(\ln{\Gamma})^3$ terms in the $\ln{\Gamma}$ expansion of the coefficients $b_i(\Gamma),c_i(\Gamma)$ could gradually become important. The lower validity threshold of $\Gamma_{\mathrm{th}}^{\mathrm{l}}\simeq5$ is imposed physically by the necessity for short-range order and imposed mathematically by the sign switching of the $b_0(\Gamma)$ coefficient. In particular, close to $\Gamma\simeq5.25$, the coefficient $b_0$ becomes negative while the coefficient $b_1$ remains positive which leads to an exponential blow-up of the bridge function at large distances.

As anticipated from the hard-sphere system, the OCP bridge function acts as additional short range repulsion. However, there is a small attractive part that arises from intermediate coupling ($\Gamma\simeq60$) up to strong coupling ($\Gamma=180$) near the first maximum of the pair correlation function\,\cite{OCPbri1,OCPbri2}. Within this coupling range, the OCP bridge function has a positive maximum whose magnitude monotonically increases from zero up to $B_{\mathrm{OCP}}\simeq0.6$ and whose position monotonically shifts from $r=1.996\,d$ up to $r=1.729\,d$. This OCP feature does not conform to the universality bridge function ansatz that was formulated by Rosenfeld and Ashcroft\,\cite{OCPbri7}. The implication is that there is space for further improvement over powerful integral theory approaches that are based on hard-sphere bridge functions such as the variational modified hypernetted-chain approximation (VMHNC)\,\cite{OCPbri8}.

\begin{figure}[!t]
        \centering
        \includegraphics[width=2.8in]{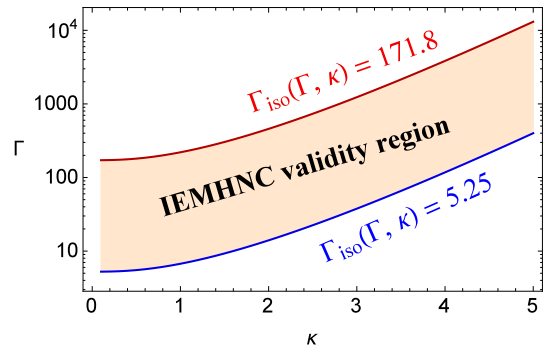}
\caption{(Color online) The validity region of the IEMHNC approximation for the YOCP system in the $(\log{\Gamma},\kappa)$ phase diagram. The region is determined by the isomorph transformation of the validity range of OCP bridge function with the upper curve roughly corresponding to the liquid-solid phase transition (the melting curve $\Gamma_{\mathrm{m}}(\kappa)$ cannot be exactly isomorph invariant due to the different configurations of the liquid and crystal phases of the YOCP\,\cite{Roskil0}) and the lower curve roughly corresponding to the onset of short-range order.}\label{validityfigure}
\end{figure}

\subsection{Bridge function for the YOCP}\label{theoryBridgeYOCP}

\noindent The proposed isomorph-based empirically modified hypernetted chain approximation (IEMHNC) utilizes the isomorph mapping to transform the simulation-extracted OCP bridge function $B_{\mathrm{OCP}}(r,\Gamma)$ into the YOCP bridge function $B_{\mathrm{YOCP}}(r,\Gamma,\kappa)$. The correspondence reads as
\begin{equation}
B_{\mathrm{YOCP}}[r/d,\Gamma,\kappa]=B_{\mathrm{OCP}}[r/d,\Gamma_{\mathrm{iso}}(\Gamma,\kappa)]\,.\label{eq-bridge7}
\end{equation}
The underlying ansatz is that the YOCP bridge function, when expressed in Wigner-Seitz units, remains completely invariant while traversing any isomorph curve. Its formulation was based on the following observations: \textbf{(i)} The application of the traditional HNC approximation to YOCP systems leads to pair correlation functions that are approximately invariant along the isomorph curves with the exception of the first peak vicinity. This implies that the addition of a completely invariant bridge function will still lead to approximately invariant pair correlation functions as observed in MD simulations\,\cite{Roskil6}, see also figures \ref{invariancefigure1},\ref{invariancefigure2}. \textbf{(ii)} A completely invariant bridge function automatically conforms to the zero-separation bridge function freezing criterion\,\cite{YOCPbr1}, known to be successful for the YOCP\,\cite{YOCPbr2}. \textbf{(iii)} Isomorph curves vary continuously up to and also including the OCP limit\,\cite{Roskil6}, in spite of the thermodynamic (but not structural) divergencies.

The validity range of the IEMHNC approximation is mainly dictated by the validity range of the OCP bridge function, which is given by $5.25\leq\Gamma_{\mathrm{OCP}}\leq171.8$. The relevant region of the YOCP phase diagram can be easily determined by applying the isomorph transform and corresponds to the region bounded by the $\Gamma_{\mathrm{iso}}(\Gamma,\kappa)=5.25$, $\Gamma_{\mathrm{iso}}(\Gamma,\kappa)=171.8$ curves, see figure \ref{validityfigure} for an illustration.

It is evident that, provided that the invariance ansatz holds, the IEMHNC accuracy depends on the accuracy of the empirical input, \emph{i.e.} on the respective accuracies of the isomorph transform and the OCP bridge function, unless some random cancellation of errors occurs. These should exhibit some fluctuations for different $(\Gamma,\kappa)$ pairs. Overall, the constructed $B_{\mathrm{YOCP}}(r/d,\Gamma,\kappa)$ is expected to be highly accurate since the YOCP is R-simple across the liquid state with very high Pearson correlation coefficients nearly approaching unity and since the insertion of the analytic bridge function of Eqs.(\ref{eq-bridge1}-\ref{eq-bridge6}) in the exact OZ closure has been demonstrated to lead to a highly accurate description of the structural and thermodynamic properties of the OCP\,\cite{OCPbri1}.

This is not the first time that the OCP bridge function has been employed as an empirical basis for the construction of the YOCP bridge function. A different functional dependence was earlier suggested\,\cite{EMHNCre}. It was based on the ad-hoc assumption that the screening parameter dependence is separable, which allowed for its determination through trial and error comparison with MD simulations. We shall refer to this approach as empirically modified hypernetted-chain approximation (EMHNC). The identified correspondence reads as\,\cite{EMHNCre}
\begin{equation}
B_{\mathrm{YOCP}}(r/d,\Gamma,\kappa)=\exp{\left(-\frac{\kappa^2}{4}\right)}B_{\mathrm{OCP}}(r/d,\Gamma)\,.\label{eq-bridge8}
\end{equation}
A detailed comparison between the IEMHNC and the EMHNC approaches will be reported in the following section. Below we shall briefly compare the two empirical bridge functions but also infer how bridge function deviations lead to structural deviations.

\begin{figure}[!ht]
        \centering\lineskip=-9pt
        \subfloat{\includegraphics[width=3.0in]{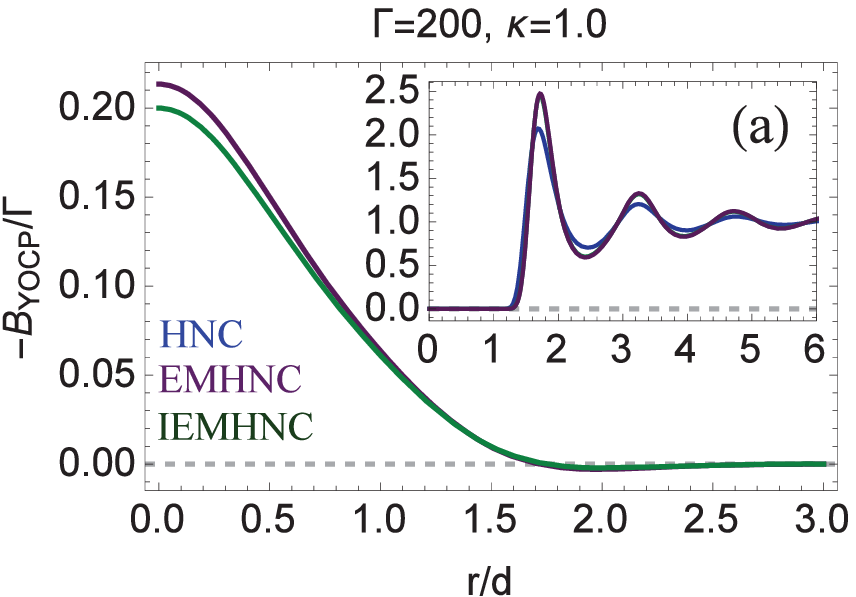}}\\
        \subfloat{\includegraphics[width=3.0in]{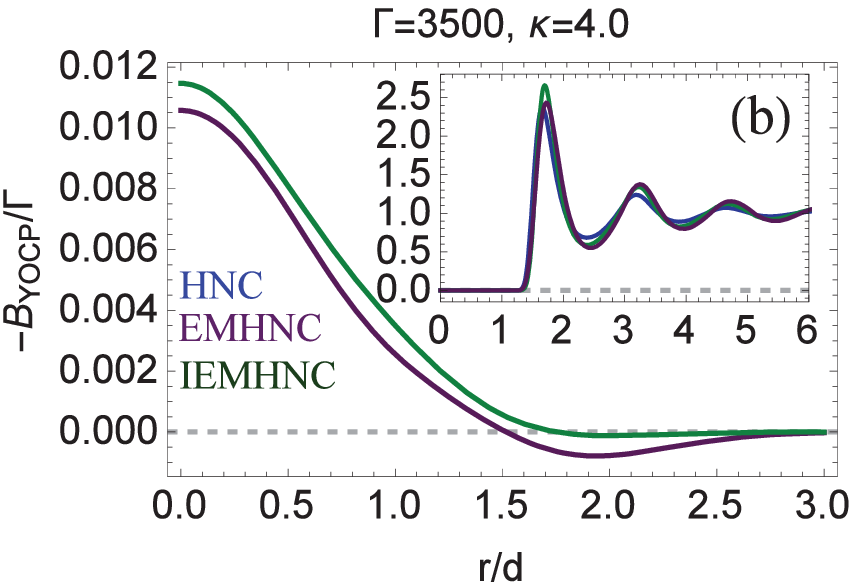}}\\
        \subfloat{\includegraphics[width=3.0in]{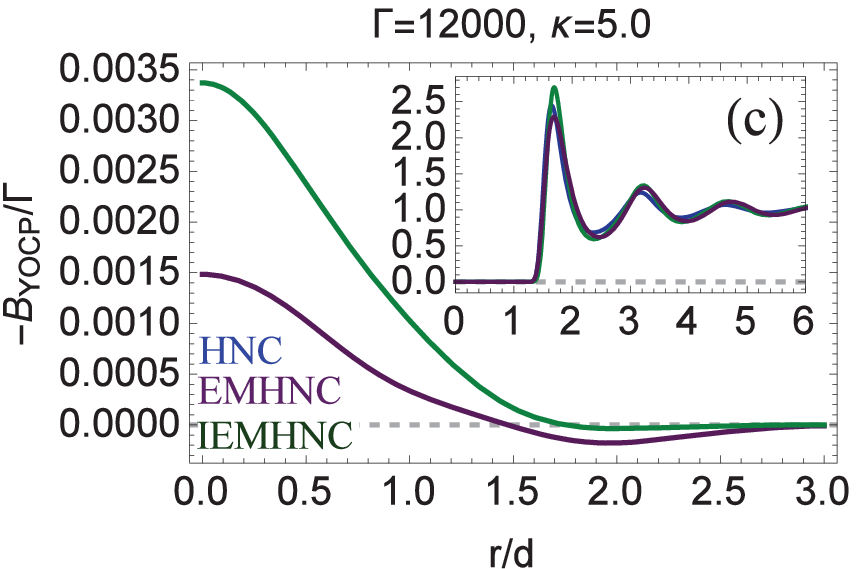}}
\caption{(Color online) The YOCP bridge functions and pair correlation functions within the HNC, EMHNC and IEMHNC approaches in three state points. (a) $\Gamma=200,\,\kappa=1.0$. The minor $B(r)$ deviations between EMHNC, IEMHNC are restricted inside the hard-core region. The resulting $g_{\mathrm{EMHNC}}(r)$ and $g_{\mathrm{IEMHNC}}(r)$ nearly overlap, they overshoot the HNC result in the first maximum vicinity, since $B(r)$ acts as an extra repulsion. (b) $\Gamma=3500,\,\kappa=4.0$. The differences extend in the whole $0\leq{r}\leq3d$ range where $B(r)$ is not fully damped. The resulting $g_{\mathrm{EMHNC}}(r)$ and $g_{\mathrm{IEMHNC}}(r)$ strongly deviate, with the EMHNC result becoming HNC-like due to the elongated positive $B(r)$ region. (c) $\Gamma=12000,\,\kappa=5.0$. The differences are further enhanced. The resulting $g_{\mathrm{EMHNC}}(r)$ and $g_{\mathrm{IEMHNC}}(r)$ diverge even more, with the HNC result overshooting the EMHNC result in the first maximum vicinity, since the EMHNC $B(r)$ now acts as an effective attraction.}\label{bridgecomparison}
\end{figure}

It is more informative to compare the bridge functions of Eqs.(\ref{eq-bridge7},\ref{eq-bridge8}) for different screening parameters, since the trends in the respective deviations are rather insensitive to the coupling parameter. In the case of very weak screening, $0<\kappa\lesssim0.7$, the two bridge functions nearly overlap in the whole range. In the case of weak screening, $0.7\lesssim\kappa\lesssim1.5$, deviations emerge at very short distances but are confined within the hard-core region, which implies a negligible effect on structural properties. In the case of intermediate screening, $1.5\lesssim\kappa\lesssim2.8$, the deviations reach the first coordination shell, which should lead to observable structural differences. In the case of strong screening, $2.8\lesssim\kappa\lesssim4.0$, the growing deviations extend in the whole range where the bridge function is non-zero. Finally, in the case of very strong screening, $\kappa\gtrsim4.0$, the EMHNC bridge function is visibly over-damped at very short distances compared to the IEMHNC bridge function and becomes positive across the whole first coordination shell, which should drastically effect the resulting correlations. In particular, this extended effective attraction suggests that the EMHNC pair correlation function should be significantly lower than the IEMHNC pair correlation function in the vicinity of the pronounced first maximum. Representative cases are shown in figure \ref{bridgecomparison}.

\section{Results}\label{numerical}

\subsection{Numerical scheme}\label{numericalscheme}

\noindent We have numerically solved Eqs.(\ref{OZequation},\ref{OZclosure}) for different bridge functions with the Picard iteration method in Fourier space\,\cite{numeri1}. Discretization was utilized allowing us to use efficient Fast Fourier Transform algorithms. Standard long range decomposition methods for the direct correlation function were followed in the weak screening regime, $0\leq\kappa<1.0$\,\cite{VMHNCre,numeri2,numeri3}. Simple mixing techniques were also employed to ensure convergence\,\cite{numeri3,numeri4}. The upper range cut-off was selected to be $R_{\mathrm{max}}=20d$. In normalized units $x=r/d$ and $q=kd$, the real space resolution was $\Delta{x}=10^{-3}$ corresponding to $20000$ points and the reciprocal space resolution was $\Delta{q}=\pi{d}/R_{\mathrm{max}}$. The convergence criterion in terms of the indirect correlation function reads as $|\gamma_{n}(q)-\gamma_{n-1}(q)|<10^{-5},\,\forall{q}$.

In an attempt to ascertain the availability of accurate initial estimates, the problem was solved in a successive manner. Namely, for each value of the screening parameter $\kappa$, solutions were initially sought for a weak coupling parameter $\Gamma$, whose value was subsequently increased in small increments with the previous outcome serving as the initial guess of the following calculation cycle.

The Picard method was preferred over the Newton-Raphson method\,\cite{numeri5}, owing to numerical problems concerning the inversion of ill-conditioned matrices which can often emerge in the weak screening regime. It was consistently observed that, in spite of the fact that the Picard method requires more iterations, the computational speed was similar. The supply of accurate initial guesses was able to remedy the well-known shortcomings of this simple method and more advanced techniques were not pursued\,\cite{numeri5,numeri6}. We emphasize that, since the IEMHNC and EMHNC approaches neither prescribe a functional relation for $B(r)$ nor contain adjustable parameters, they can be readily implemented in existing HNC algorithms.

\subsection{Pair correlation functions}\label{numericalcorrelation}

\begin{figure}
        \centering\lineskip=-8pt
        \subfloat{\includegraphics[width=2.97in]{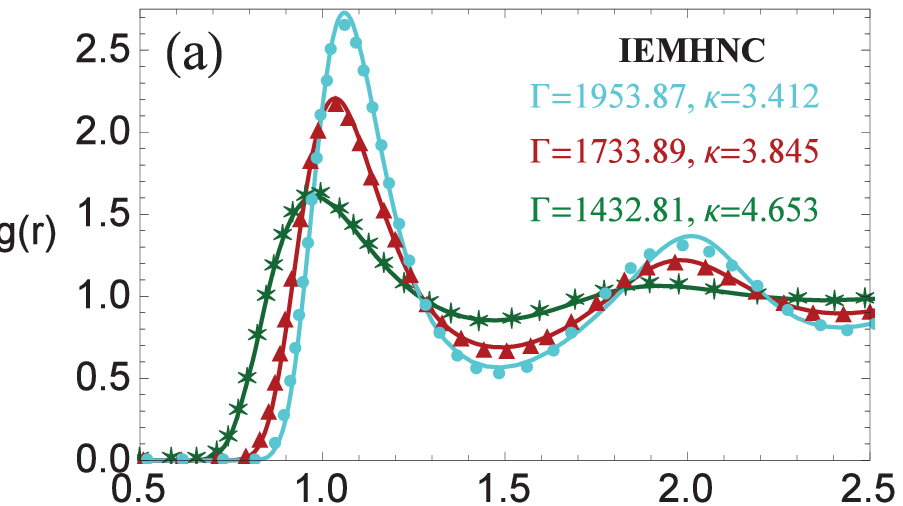}}\\
        \subfloat{\includegraphics[width=2.97in]{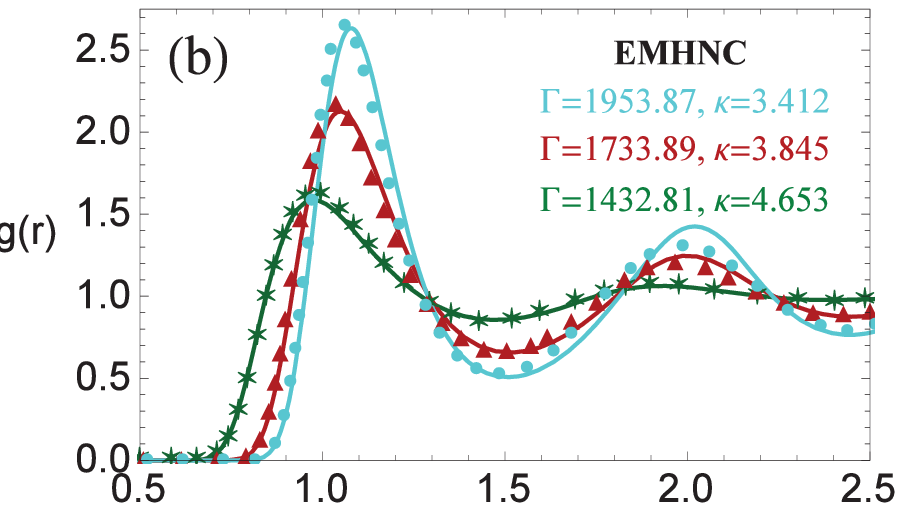}}\\
        \subfloat{\includegraphics[width=2.97in]{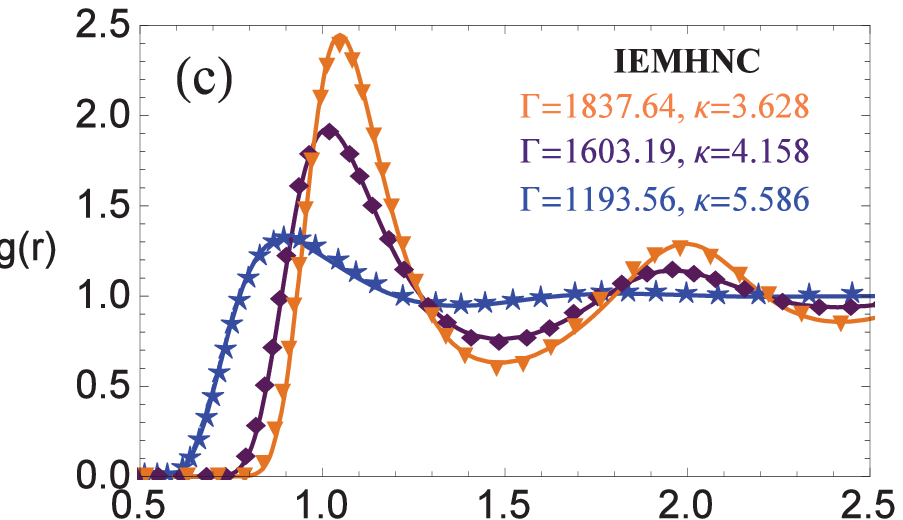}}\\
        \subfloat{\includegraphics[width=2.97in]{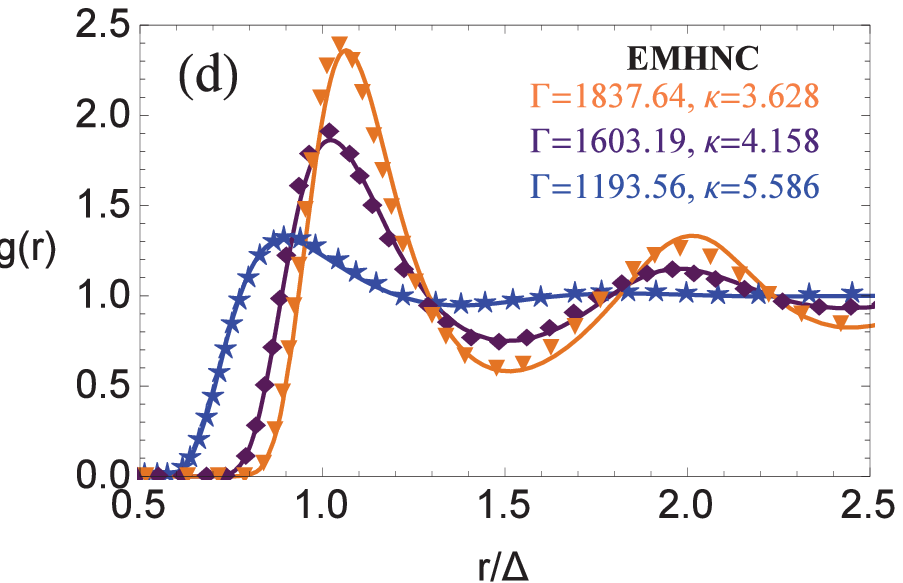}}
\caption{(Color online) Pair correlation functions resulting from MD simulations (discrete points) and integral theory methods (solid lines), namely (a,c) the IEMHNC approach and (b,d) the EMHNC approach. Compare sub-figures (a) with (b), (c) with (d). The MD results have been adopted from Ref.\cite{Roskil6} and have been sparsely sampled in order to facilitate comparison. In the $(\widetilde{t},\widetilde{n})$ phase diagram, the state points of interest correspond to the isotherm $\widetilde{t}=1.5\times10^{-4}$ and $\widetilde{n}=1.37\times10^{-3},\,2.37\times10^{-3},\,3.32\times10^{-3},\,4.20\times10^{-3},\,5.00\times10^{-3},\,6.01\times10^{-3}$ where the normalized temperature is defined by $\widetilde{t}=T\lambda/Q^2$ and the normalized density by $\widetilde{n}=n\lambda^3$. The mapping between the $(\widetilde{t},\widetilde{n})$ and $(\Gamma,\kappa)$ phase variables reads as $\kappa^3=3/(4\pi\widetilde{n})$, $\Gamma^3=(4\pi\widetilde{n})/(3\widetilde{t}^3)$. The distance has been normalized to the cubic mean inter-particle distance $\Delta=n^{-1/3}$.}\label{paircorrelationcomparison}
\end{figure}

\noindent The IEMHNC accuracy is so high that graphical comparisons with published simulation-generated pair correlation functions might not be informative, especially in case of weak screening parameters. Some characteristic examples for YOCP state points that belong to the strong screening range are illustrated in figure \ref{paircorrelationcomparison}; the IEMHNC approach well reproduces the MD simulations (see \ref{paircorrelationcomparison}a and \ref{paircorrelationcomparison}c), whereas the EMHNC approach exhibits observable deviations predominantly close to the first and second peaks (see \ref{paircorrelationcomparison}b and \ref{paircorrelationcomparison}d).

In recent comprehensive LD simulations of the weakly screened YOCP\,\cite{Bonitz2}, some key functional properties of the pair correlation function were extracted. These concerned the distance at which $g(r)=1/2$ as a measure of the correlation void, the magnitudes and positions of the first maximum, first nonzero minimum and second maximum as measures of the first two coordination shells. In the LD simulations, three screening parameters were probed $(\kappa=0,\,1,\,2)$ and the coupling parameter varied from unity up to the crystallization vicinity. As aforementioned, for such weak screening, the deviations between the IEMHNC and the EMHNC approximations are expected to be very small. Nevertheless, these structural data allow for the quantification of the accuracy of theoretical approaches better than thermodynamic properties, where integral operators can play a smoothing role. Tabulations of these functional characteristics as resulting from the LD simulations and from different integral theory methods (IEMHNC, EMHNC and HNC as reference) have been compiled in the supplemental material.

Here we shall mainly focus on the correlation void and the magnitude of the first maximum. Overall, the IEMHNC and EMHNC approaches are almost indistinguishable from simulations. \textbf{(i)} In the case of $\kappa=1.0$, the IEMHNC and EMHNC yield nearly identical results with a drastic improvement over the HNC, especially concerning the magnitude of the first maximum. The IEMHNC leads to more accurate first maximum predictions, while the EMHNC leads to more accurate correlation void predictions. The deviations from LD simulations are $\lesssim1\%$. \textbf{(ii)} In the case of $\kappa=2.0$, the IEMHNC becomes noticeably superior in the range $\Gamma/\Gamma_{\mathrm{m}}\gtrsim0.4$. This primarily concerns the first maximum magnitude, with IEMHNC deviations never exceeding $1.2\%$ and EMHNC deviations reaching $4.1\%$. \textbf{(iii)} Compared to the OCP, as the screening parameter increases, the IEMHNC accuracy remains nearly the same as far as the seven studied functional $g(r)$ properties are concerned. This does not apply for the EMHNC and indirectly supports the ansatz of isomorph invariant bridge functions behind the IEMHNC.

\subsection{Reduced excess internal energy and pressure}\label{numericalthermodynamic}

\noindent Extensive MD results of the \emph{reduced excess internal energy} owing to particle-particle interactions, given by
\begin{equation*}
u_{\mathrm{ex}}^{\mathrm{pp}}(\Gamma,\kappa)=\frac{1}{2}\beta{n}\int\,u(r)g(r)d^3r\,,
\end{equation*}
have been reported in the literature\,\cite{Hamagu1,Hamagu2,Hamagu3}. The YOCP results span from the weak coupling regime up to the liquid-solid transition and from the unscreened Coulomb limit up to the strong screening range. Systematic tabulations of $u_{\mathrm{ex}}^{\mathrm{pp}}$ stemming from MD simulations and different integral theory methods (IEMHNC, EMHNC, HNC and SMSA) are available in the supplemental material.

The primary conclusions drawn from the comparison are summarized in what follows: \textbf{(i)} The IEMHNC approach exhibits an excellent agreement with simulations. For $\kappa=\{0.2,\,0.4,\,0.6,\,0.8,\,1.0,\,1.2,\,1.4\}$, MD results are reproduced well within their published fluctuation levels. For higher screening parameters, within the dense fluid region (loosely defined by $0.25\lesssim\Gamma/\Gamma_{\mathrm{m}}\lesssim1$ or equivalently $40\lesssim\Gamma_{\mathrm{iso}}^{\mathrm{OCP}}\lesssim171.8$) the relative deviations from MD results are always below $\lesssim0.5\%$. \textbf{(ii)} The IEMHNC results are nearly identical with the EMHNC results for $\kappa\lesssim1.4$. For stronger screening, the IEMHNC approach performs consistently better than the EMHNC when $\Gamma/\Gamma_{\mathrm{m}}\gtrsim0.2$, whereas the EMHNC approach performs consistently better than the IEMHNC when $\Gamma/\Gamma_{\mathrm{m}}\lesssim0.1$. For the strongest screening investigated ($\kappa=5.0$), the EMHNC deviations from MD results even reach $\sim4.5\%$ in the dense fluid region. \textbf{(iii)} For the OCP, the IEMHNC deviations from MD are abruptly decreasing with the coupling parameter near $\Gamma/\Gamma_{\mathrm{m}}\simeq0.1$, remain nearly constant up to $\Gamma/\Gamma_{\mathrm{m}}\simeq0.8$ and slowly increase up to $\Gamma\simeq\Gamma_{\mathrm{m}}$. The above $\Gamma-$pattern roughly persists for the YOCP regardless of the screening parameter, which indirectly supports the ansatz of isomorph invariant bridge functions.

Rather limited MC results are available for the \emph{reduced excess pressure} owing to particle-particle interactions, as obtained from the virial expression\,\cite{Meijer1,DRYpape}
\begin{equation*}
p_{\mathrm{ex}}^{\mathrm{pp}}(\Gamma,\kappa)=-\frac{1}{6}\beta{n}\int\,r\frac{du}{dr}g(r)d^3r\,.
\end{equation*}
We should first elaborate on the fact that the use of the energy route to the equation of state for comparison between theory and simulations can be restrictive. For the YOCP, the latter thermodynamic path is described by
\begin{equation*}
p_{\mathrm{ex}}^{\mathrm{pp}}(\Gamma,\kappa)=-\frac{\kappa}{3}\frac{\partial{f}_{\mathrm{ex}}^{\mathrm{pp}}(\Gamma,\kappa)}{\partial\kappa}+\frac{u_{\mathrm{ex}}^{\mathrm{pp}}(\Gamma,\kappa)}{3}\,,
\end{equation*}
with $f_{\mathrm{ex}}^{\mathrm{pp}}(\Gamma,\kappa)=\int_0^{\Gamma}\,[u_{\mathrm{ex}}^{\mathrm{pp}}(\Gamma^{\prime},\kappa)/\Gamma^{\prime}]d\Gamma^{\prime}$ the reduced excess free energy. The integration of a discrete function is required, which would inevitably introduce interpolation or numerical quadrature errors since it is highly unlikely that the available simulation data are sufficiently dense. These errors would hinder meaningful comparison with highly accurate approaches such as the IEMHNC.

Tabulated $p_{\mathrm{ex}}^{\mathrm{pp}}$ data resulting from MC simulations and from different integral theory approaches (IEMHNC, EMHNC, HNC, SMSA, the Discretized Rogers-Young or DRY method) can be found in the supplemental material. The IEMHNC performance is again superior to all other approximations with the relative deviations from the MC results being mostly $<0.1\%$ and with the maximum deviation being $\simeq0.39\%$. We emphasize that the IEMHNC approach is even more accurate than the DRY method, despite the fact that the latter's assumed bridge function contains an adjustable parameter that is determined by enforcing thermodynamic consistency\,\cite{DRYpape,approx6}.

\subsection{Thermodynamic consistency}\label{numericalconsistency}

\noindent As an additional test, we have examined the IEMHNC thermodynamic consistency between the statistical and the virial paths to the compressibility of the YOCP. The \emph{statistical route} to the reduced excess inverse isothermal compressibility due to particle presence (where the contribution of particle-background interactions is added to the contribution of particle-particle interactions) leads to
\begin{equation*}
\mu_{\mathrm{stat}}^{\mathrm{p}}(\Gamma,\kappa)=-n\int\,\left[c(r)+\beta{u}(r)\right]d^3r\,.
\end{equation*}
We point out that, courtesy of the exact asymptotic limit $c(r)\to-\beta{u}(r)$, the truncation of the integration range at $r=20d$ should have a negligible effect. On the other hand, the \emph{virial route} leads to
\begin{equation*}
\mu_{\mathrm{vir}}^{\mathrm{p}}(\Gamma,\kappa)=p_{\mathrm{ex}}^{\mathrm{p}}(\Gamma,\kappa)+\frac{\Gamma}{3}\frac{\partial{p}_{\mathrm{ex}}^{\mathrm{p}}(\Gamma,\kappa)}{\partial\Gamma}-\frac{\kappa}{3}\frac{\partial{p}_{\mathrm{ex}}^{\mathrm{p}}(\Gamma,\kappa)}{\partial\kappa}\,.
\end{equation*}
with $p_{\mathrm{ex}}^{\mathrm{p}}=-(1/6)\beta{n}\int\,r\left(du/dr\right)\left[g(r)-1\right]d^3r$ denoting the reduced excess pressure due to the particle presence. We point out that, courtesy of the exact asymptotic limit $g(r)\to1$, the $r=20d$ truncation should again have a negligible effect. The required first-order derivatives with respect to $(\Gamma,\kappa)$ have been computed with the central difference method for $\Delta\Gamma=\pm0.001\Gamma$ and $\Delta\kappa=\pm0.001\kappa$, which is a tedious procedure since the determination of $\mu_{\mathrm{vir}}^{\mathrm{p}}$ at each state point requires the numerical solutions of the integral equation system at five state points.

\begin{figure}[!t]
        \centering\lineskip=-18pt
        \subfloat{\includegraphics[width=2.9in]{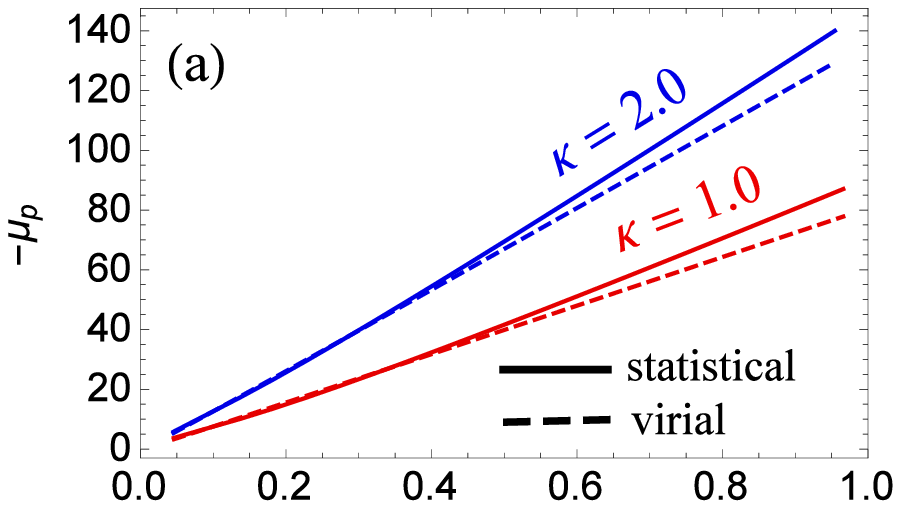}}\\
        \subfloat{\includegraphics[width=2.9in]{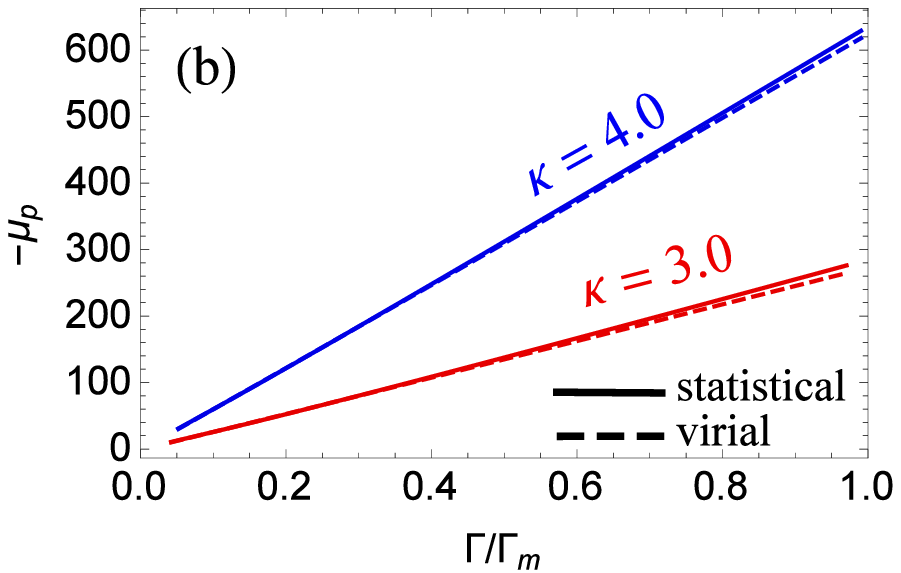}}
\caption{(Color online) Test of the thermodynamic consistency of the IEMHNC approximation. The reduced excess inverse isothermal compressibility due to the particle presence resulting from the statistical and virial routes as a function of coupling parameter. Results for integer values of the screening parameter; (a) $\kappa=1.0,\,\kappa=2.0$, (b) $\kappa=3.0,\,\kappa=4.0$.}\label{consistencyfigure}
\end{figure}

Exhaustive tabulations of $(\mu_{\mathrm{stat}}^{\mathrm{p}},\mu_{\mathrm{vir}}^{\mathrm{p}})$ stemming from the IEMHNC and EMHNC approximations are available in the supplemental material for $15$ screening parameters $\kappa\in[0,5]$ and coupling parameters $\Gamma/\Gamma_{\mathrm{m}}\in[0.1,1]$. The IEMHNC statistical - virial deviations are maximum near the OCP crystallization point, where they reach $\sim15\%$. Furthermore, for a constant $\Gamma/\Gamma_{\mathrm{m}}$, the IEMHNC deviations monotonically decrease as the screening parameter increases (see figure \ref{consistencyfigure}), for instance when $\kappa=5.0$ they are always $\lesssim0.6\%$ regardless of the coupling parameter. Finally, for any screening parameter, the IEMHNC approach performs systematically better than the EMHNC within the dense fluid region. Overall, a very high degree of thermodynamic consistency has been verified across the YOCP phase space. This IEMHNC feature is rather remarkable, given the absence of adjustable parameters.

\subsection{Approximate isomorph invariance of the YOCP static pair correlations}\label{numericalinvariance}

\noindent The bridge function invariance ansatz of the IEMHNC approach states that the reduced unit YOCP bridge functions remain completely unchanged while traversing isomorph curves. In the absence of bridge functions, as observed from the HNC approach, YOCP structural properties exhibit an approximate invariance along isomorph lines. It is clear that the addition of invariant bridge functions to the integral theory framework will result to more invariant static correlations. Nevertheless, as observed in computer simulations, the invariance of the pair correlation functions and structure factors remains approximate within the IEMHNC approach.

Since YOCP liquids belong to the class of R-simple systems, the static structure factor $S(q)$ should be an approximate isomorph invariant when expressed in Wigner-Seitz units $q=kd$\,\cite{Roskil0}. However, this does not apply for excess thermodynamic quantities that involve first or second order volume derivatives\,\cite{Roskil0,Roskil3}, such as the excess pressure $P_{\mathrm{ex}}=-\left(\partial{F}_{ex}/\partial{V}\right)_{T,N}$ and the excess inverse isothermal compressibility $\mu_{\mathrm{T}}=-V\left(\partial{P}_{\mathrm{ex}}/\partial{V}\right)_{T,N}$. The above statements seem to contradict each other in view of the well-known connection between the long-wavelength limit of the static structure factor and the isothermal compressibility\,\cite{bookre1,bookre2}. The approximate nature of isomorph invariance as well as the gradual increase of the virial-potential energy correlation coefficient in the vicinity of the melting line resolve this issue. In particular, as the isomorph lines approach the nearly parallel melting line, the correlation coefficient approaches unity and the structure factors approach exact invariance which is thermodynamically allowed owing to the very small values of the isothermal compressibility for any state point belonging to the isomorph, see also figure \ref{consistencyfigure}. Conversely, for isomorph lines far from the melting line, the YOCP compressibility values increase with the relative variations within the same isomorph becoming more observable, whereas the correlation coefficient decreases and deviations arise between isomorphic structure factors.

\begin{figure}
        \centering\lineskip=-10pt
        \subfloat{\includegraphics[width=2.97in]{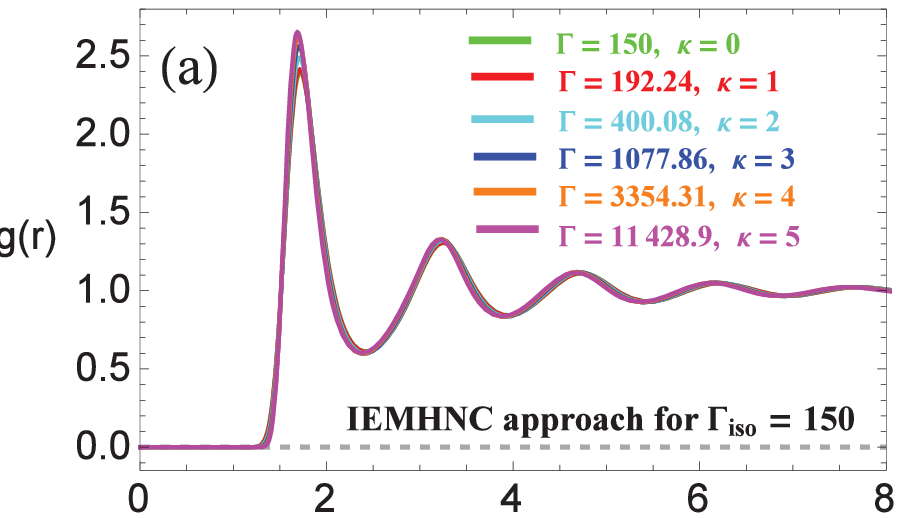}}\\
        \subfloat{\includegraphics[width=2.97in]{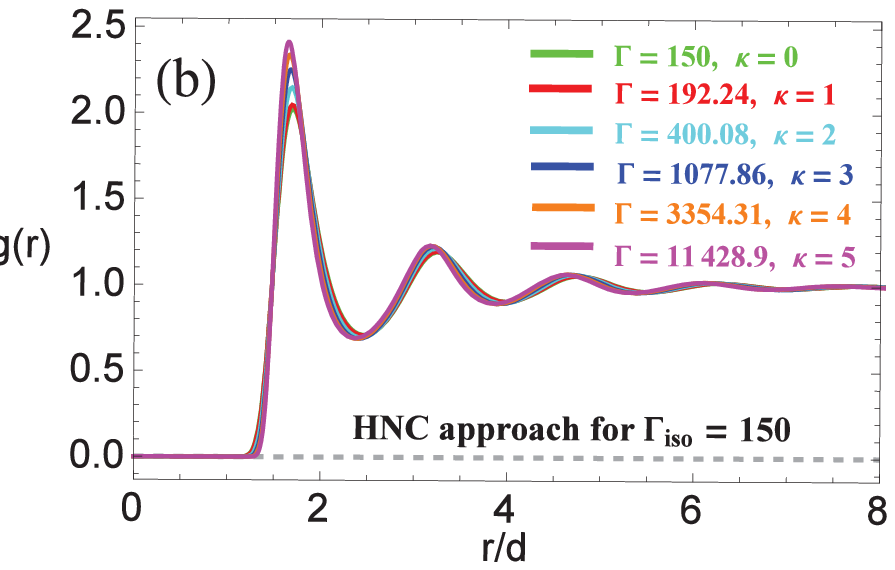}}\\
        \subfloat{\includegraphics[width=2.97in]{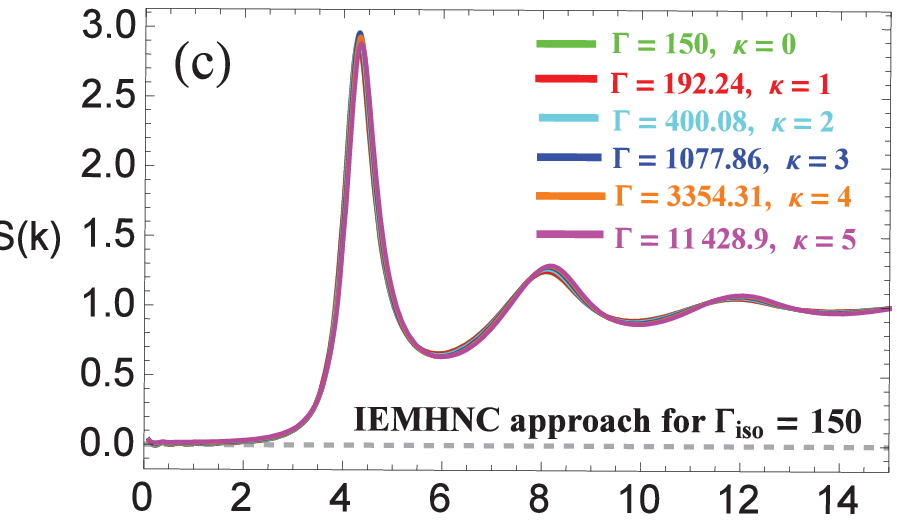}}\\
        \subfloat{\includegraphics[width=2.97in]{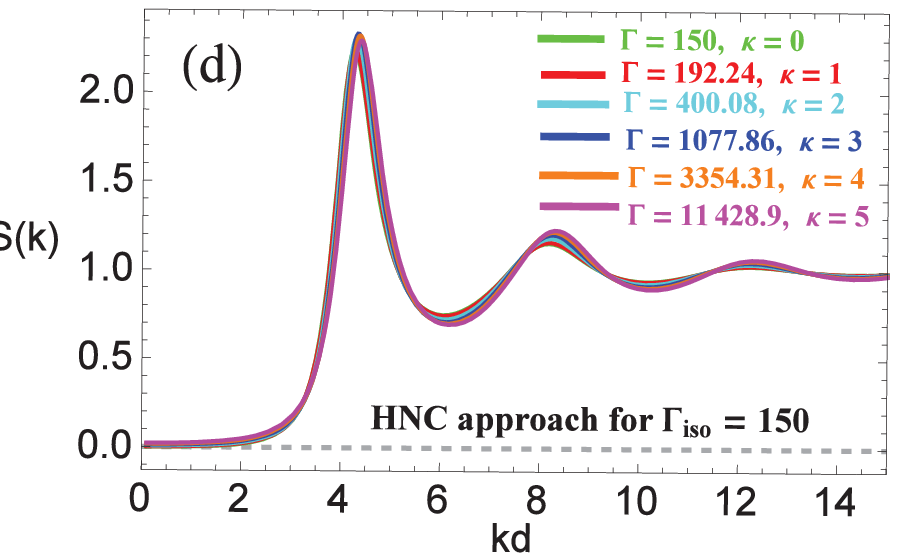}}
\caption{(Color online) The YOCP pair correlation functions and static structure factors as computed from the IEMHNC and HNC approach. Results for six YOCP state points along the isomorph $\Gamma_{\mathrm{iso}}(\Gamma,\kappa)=150$ that correspond to the normalized screening parameters $\kappa=0,\,1,\,2,\,3,\,4,\,5$. The isomorph curve lies close to the melting line that can be approximated by $\Gamma_{\mathrm{iso}}(\Gamma,\kappa)=171.8$. The addition of the IEMHNC bridge function naturally improves the isomorph invariance of both static quantities. We note that the structure factor is noticeably more invariant than the pair correlation function. The non-invariant features are nearly exclusively concentrated in the vicinity of the first maximum, in accordance with MD simulation studies of most R-simple systems\,\cite{Roskil0,Roskil7}.}\label{invariancefigure1}
\end{figure}

\begin{figure}
        \centering\lineskip=-7,5pt
        \subfloat{\includegraphics[width=2.97in]{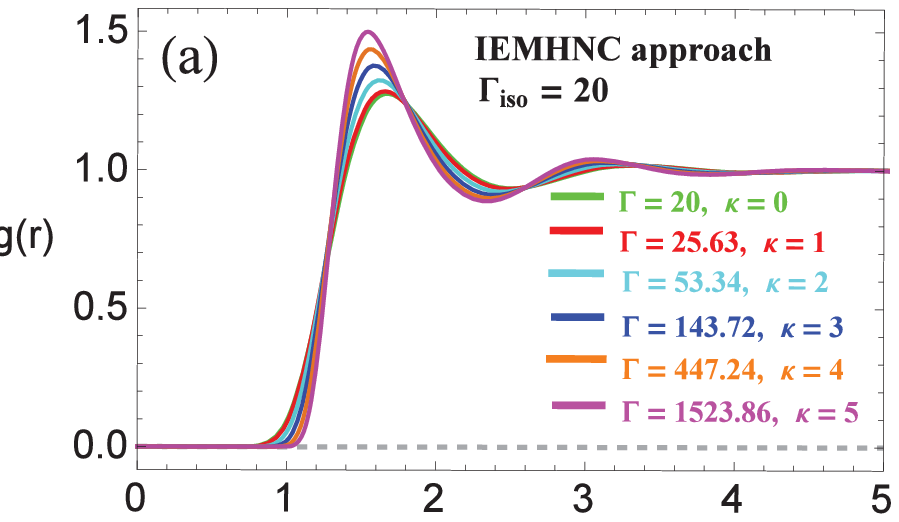}}\\
        \subfloat{\includegraphics[width=2.97in]{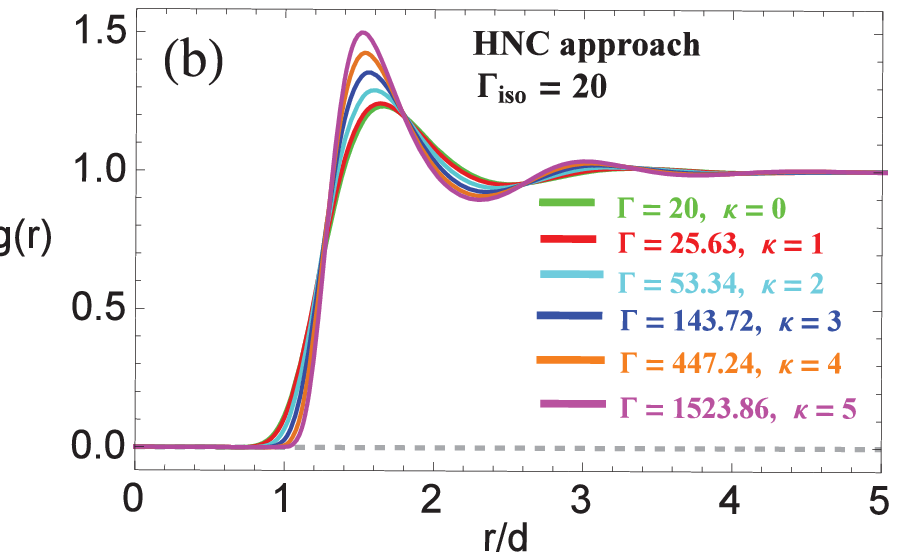}}\\
        \subfloat{\includegraphics[width=2.97in]{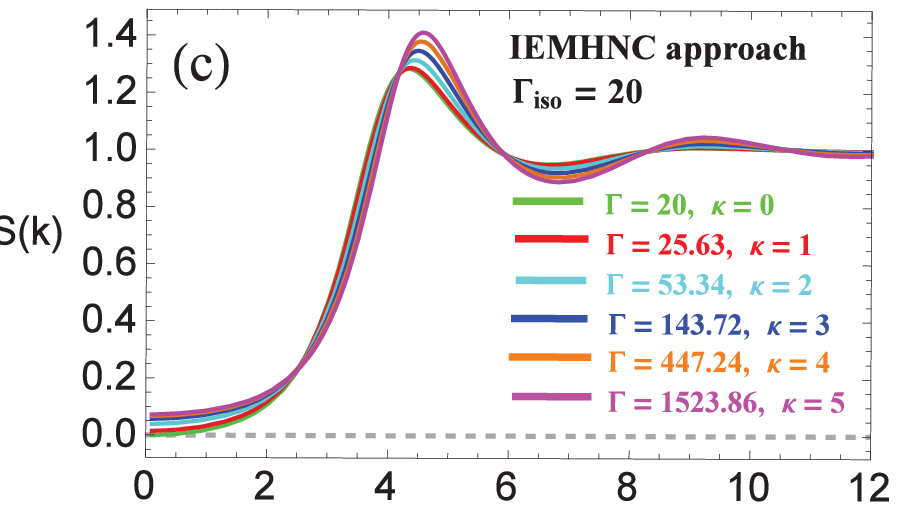}}\\
        \subfloat{\includegraphics[width=2.97in]{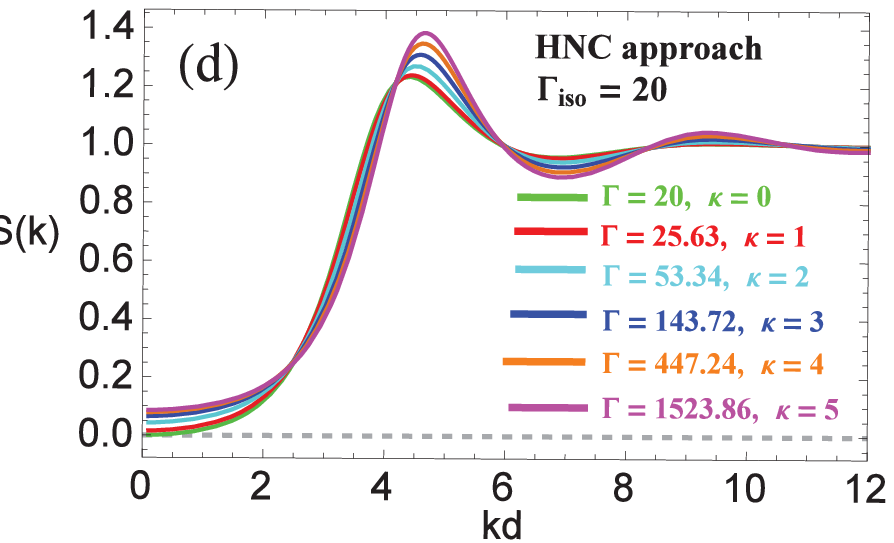}}
\caption{(Color online) The YOCP pair correlation functions and static structure factors as computed from the IEMHNC and HNC approach. Results for six YOCP state points along the isomorph $\Gamma_{\mathrm{iso}}(\Gamma,\kappa)=20$ that correspond to the normalized screening parameters $\kappa=0,\,1,\,2,\,3,\,4,\,5$. The isomorph curve lies far from the melting line that can be approximated by $\Gamma_{\mathrm{iso}}(\Gamma,\kappa)=171.8$. The addition of the IEMHNC bridge function in this regime barely improves the invariance of both static quantities and the IEMHNC results are close to the HNC results. Note that the long-wavelength behavior of the structure factor is strongly non-invariant due to the varying compressibility of the state points. The non-invariant features are again more pronounced in the vicinity of the first peak.}\label{invariancefigure2}
\end{figure}

These remarks are better illustrated through examples. In figures \ref{invariancefigure1},\ref{invariancefigure2}, the pair correlation functions and structure factors along two YOCP isomorph lines, $\Gamma_{\mathrm{iso}}(\Gamma,\kappa)=150$ and  $\Gamma_{\mathrm{iso}}(\Gamma,\kappa)=20$, are plotted as resulting from the IEMHNC and HNC approaches. In the former case, the isomorph line approaches the melting line $\Gamma_{\mathrm{iso}}(\Gamma,\kappa)=171.8$ and both static quantities are almost exactly invariant. In the latter case, the isomorph line is far from the melting line and both static quantities are barely invariant. The addition of the bridge function always improves the invariance, but this is only visible for $\Gamma_{\mathrm{iso}}(\Gamma,\kappa)=150$ due to the well-known diminishing role of the bridge function away from melting.

\section{Discussion}\label{discussion}

\subsection{Generalization of the bridge function construction scheme to arbitrary R-simple systems}\label{discgene}

\noindent Let us consider a Roskilde-simple one-component system that consists of point particles which interact with the isotropic pair potential $u(r)=(1/\epsilon)V(r/\sigma)$, where $\epsilon$ and $\sigma$ denote the characteristic energy and length scales. The thermodynamic state of the system is fully described by the dimensionless variables $\widetilde{n}=n\sigma^3$, $\widetilde{t}=k_{\mathrm{b}}T/\epsilon$. The R-simple system exhibits approximate invariance along any line belonging to the isomorph set described by $f(\widetilde{n},\widetilde{t})=\mathrm{constant}$. The a priori knowledge of the bridge function along any phase diagram curve that has a single intersection point with each member of the isomorph family of curves would allow for the straightforward extension of the bridge function in the whole phase diagram within the exact invariance ansatz.

To be more concrete, computer simulations can be carried out to extract two dimensional bridge functions $B(r/d,\widetilde{n}_{\mathrm{iso}})$ along an arbitrary isotherm $\widetilde{t}_{\mathrm{ref}}$ and the isomorph mapping $f(\widetilde{n},\widetilde{t})=f(\widetilde{n}_{\mathrm{iso}},\widetilde{t}_{\mathrm{ref}})$ can then be solved for $\widetilde{n}_{\mathrm{iso}}(\widetilde{n},\widetilde{t})$ to construct the three dimensional bridge function $B[r/d,\widetilde{n}_{\mathrm{iso}}(\widetilde{n},\widetilde{t})]$. In a similar fashion, simulations can be employed to extract two dimensional bridge functions $B(r/d,\widetilde{t}_{\mathrm{iso}})$ along an arbitrary isochore $\widetilde{n}_{\mathrm{ref}}$ and the isomorph mapping $f(\widetilde{n},\widetilde{t})=f(\widetilde{n}_{\mathrm{ref}},\widetilde{t}_{\mathrm{iso}})$ can then be solved for $\widetilde{t}_{\mathrm{iso}}(\widetilde{n},\widetilde{t})$ in order to construct the three dimensional bridge function $B[r/d,\widetilde{t}_{\mathrm{iso}}(\widetilde{n},\widetilde{t})]$.

\subsection{Extension of the IEMHNC approximation to bi-Yukawa systems}\label{discbiyu}

\noindent Repulsive bi-Yukawa pair-interactions can be utilized as model potentials of strongly coupled systems whose inter-particle interactions are characterized by two fundamental length scales. For instance, in laboratory realizations of isotropic complex plasmas, dust-dust interactions involve
a short-range characteristic length due to the polarization of the plasma background and a long-range characteristic length due to the competition between plasma ionization and absorption or recombination\,\cite{biYOCP1,biYOCP2,biYOCP3,biYOCP4}. Furthermore, in warm dense matter, effective ion-ion interactions involve a long-range characteristic length due to the polarization of the free electrons and a short-range repulsion due to the overlapping of the bound electron wavefunctions\,\cite{WDMpap1,WDMpap2,WDMpap3,WDMpap4}. Expressed in Wigner-Seitz coordinates $x=r/d$, the pair interaction energy of such bi-Yukawa one-component plasmas (biYOCP) becomes
\begin{equation*}
\beta{u}(x)=\frac{\Gamma}{x}\left[(1-\sigma)\exp{\left(-\kappa{x}\right)}+\sigma\exp{\left(-\mu\kappa{x}\right)}\right]\,,
\end{equation*}
where $(\Gamma,\kappa)$ are the dimensionless thermodynamic state variables and $(\sigma,\mu)$ are external potential parameters.

There are two important properties of dense repulsive biYOCP liquids that allow for the direct application of the IEMHNC approximation. In the limit $\kappa\to0$, the interaction energy becomes $\beta{u}(x)=\Gamma/x$ and the biYOCP collapses to the OCP. According to an additivity theorem\,\cite{Roskil2,Roskil3}, the biYOCP should be R-simple in an extensive region of its liquid phase diagram with the respective isomorph mapping described by
\begin{equation*}
\Gamma_{\mathrm{iso}}^{\mathrm{biYOCP}}(\Gamma,\kappa;\sigma,\mu)=(1-\sigma)\Gamma_{\mathrm{iso}}^{\mathrm{YOCP}}(\Gamma,\kappa)+\sigma\Gamma_{\mathrm{iso}}^{\mathrm{YOCP}}(\Gamma,\mu\kappa)
\end{equation*}
which, in view of Eq.(\ref{eq-mapping}), is equivalent to
\begin{align*}
\Gamma_{\mathrm{iso}}^{\mathrm{biYOCP}}&(\Gamma,\kappa;\sigma,\mu)=\Gamma\left\{(1-\sigma){e}^{-\alpha\kappa}\left[1+\alpha\kappa+\frac{(\alpha\kappa)^2}{2}\right]\nonumber\right.\\&\left.+\sigma{e}^{-\mu\alpha\kappa}\left[1+\mu\alpha\kappa+\frac{(\mu\alpha\kappa)^2}{2}\right]\right\}=\mathrm{const.}
\end{align*}
Combining the above, within the exact bridge function invariance ansatz, the IEMHNC bridge function for the biYOCP system can be constructed from the expression $B_{\mathrm{biYOCP}}[x,\Gamma,\kappa;\sigma,\mu]=B_{\mathrm{OCP}}[x,\Gamma_{\mathrm{iso}}^{\mathrm{biYOCP}}(\Gamma,\kappa;\sigma,\mu)]$.

\subsection{Possible improvements of the IEMHNC approximation}\label{discimprov}

\noindent The IEMHNC approximation is characterized by a very high level of accuracy in the entire liquid YOCP phase diagram without the implementation of any adjustable parameters. Consequently, potential improvement schemes will not be investigated in a detailed manner. Nevertheless, it is still worth mentioning such possibilities in view of extensions to the other R-simple systems, where the IEMHNC approach might be less accurate. At this point, we should emphasize that the success of the IEMHNC approach for dense YOCP liquids merely indicates that the YOCP bridge functions are approximate isomorph invariants. It is not possible to quantify the level of invariance, because of the weak sensitivity of the pair correlation function on the bridge function\,\cite{lastcit}.

We should first reiterate that the simulation-extracted OCP bridge function is not exact; a relatively small number of coupling parameters was analyzed by MC simulations\,\cite{OCPbri1}, the short-range extrapolation procedure has been criticized\,\cite{OCPbri5,OCPbri6} and the supercooled regime was not considered. In section \ref{numerical}, it was observed that the IEMHNC deviations from simulations and the IEMHNC thermodynamic consistency do not significantly deteriorate as the screening parameter increases indicating that discrepancies in the OCP bridge function and not in the isomorph mapping constitute the main source of errors. The above strongly suggest that the systematic extraction of more accurate OCP bridge functions from dedicated simulations and their subsequent parameterization are rather imperative future tasks.

Different improvement schemes can be formulated by inserting adjustable parameters to the YOCP bridge function that can be uniquely determined by comparing with MD simulation results, by enforcing thermodynamic consistency\,\cite{bookre3} or imposing empirical freezing rules\,\cite{YOCPbr1}. \textbf{(i)} The quantity $\alpha=\Delta/d$ in the isomorph mapping could be treated as a function of the screening parameter\,\cite{Roskil6}, \emph{i.e.} $\alpha(\kappa)=f(\kappa)(\Delta/d)$ where $f(\kappa)$ should obtain values close to unity. Notice that this modification does not violate the ansatz of exact bridge function invariance. \textbf{(ii)} Taking the EMHNC bridge function into account\,\cite{EMHNCre}, the YOCP bridge function could be factorized in the following manner $B_{\mathrm{YOCP}}[r,\Gamma,\kappa]=f(\kappa)B_{\mathrm{YOCP}}^{\mathrm{IEMHNC}}[r,\Gamma,\kappa]$ with $f(\kappa)$ an unknown function, where it is assumed that non-invariant effects simply lead to a re-scaling of the bridge function. \textbf{(iii)} Taking the VMHNC bridge function into consideration\,\cite{VMHNCre}, the YOCP bridge function could be decomposed into an invariant part and a hard-core part with the unknown effective packing fraction $\eta_{\mathrm{eff}}$ depending on both the coupling and screening parameters, where it is assumed that non-invariant effects modify the functional form of the bridge function.

The availability of simulation extracted YOCP bridge functions could provide guidance to various improvement schemes. More important, it would allow for a rigorous test of the validity of the ansatz of exact bridge function invariance, which constitutes the basis of the proposed IEMHNC approach. Unfortunately, preliminary investigations of this kind were discontinued shortly after their initiation\,\cite{Cailol2}.

\section{Summary}\label{outro}

\noindent An analytical expression has been proposed for the bridge function of Yukawa liquids that is obtained by applying a configurational adiabat transform (isomorph mapping) to available simulation-extracted bridge functions for strongly coupled Coulomb systems. The underlying ansatz postulates that the Yukawa bridge function (when expressed in Wigner-Seitz units) remains exactly invariant while traversing any isomorph curve up to and including the unscreened limit. Introduction of the bridge function to the standard integral theory formalism leads to the isomorph-based empirically modified hypernetted-chain approximation. To our knowledge, this is the first time that isomorph invariance have been utilized as one of the building blocks of a theoretical approach.

The resulting structural characteristics (pair correlation functions) and thermodynamic properties (excess internal energy, excess pressure) revealed an excellent agreement with computer simulations in the entire dense liquid region of the Yukawa phase diagram. Systematic comparisons were carried out with other integral theory methods, which revealed the superiority of the proposed approach in the strongly coupled regime. The approximate thermodynamic consistency of the virial and statistical routes to the compressibility was demonstrated from the bare Coulomb regime up to the strong screening range. The bridge function construction scheme was generalized for arbitrary Roskilde-simple one-component systems and extended to dense bi-Yukawa plasmas. Possible improvement schemes were discussed. The need for simulation-extracted Coulomb and Yukawa bridge functions was also pointed out.

\section*{Supplementary material}

\noindent See the Supplemental Material for quantitative comparisons of the results of the isomorph-based empirically modified hypernetted-chain approximation with the results of computer simulations and various integral theory approaches. The material contains extensive tabulations of key functional properties of the YOCP pair correlation functions and tabulations of some basic YOCP thermodynamic quantities (internal energy, pressure, statistical and virial compressibility).

\section*{Acknowledgments}

\noindent The authors would like to acknowledge the financial support of the Swedish National Space Agency. FLC would like to thank Torben Ott and Michael Bonitz for providing unpublished errata to the supplemental material of Ref.\cite{Bonitz2}.

\end{document}